\documentclass[prl, twocolumn, showpacs, superscriptaddress]{revtex4-2}
\usepackage{amsmath, amssymb}
\usepackage{graphicx}
\renewcommand{\section}[1]{{\par\it #1.---}\ignorespaces}

\begin{document}
\title{An explicit evolution from N{\'e}el to striped antiferromagnetic states in the spin-1/2 $J_{1}$-$J_{2}$ Heisenberg model on the square lattice }
\author{Yun-Tong Yang}
\affiliation{School of Physical Science and Technology $\&$ Lanzhou Center for Theoretical Physics, Lanzhou University, Lanzhou 730000, China}
\affiliation{Key Laboratory of Quantum Theory and Applications of MoE $\&$ Key Laboratory of Theoretical Physics of Gansu Province, Lanzhou University, Lanzhou 730000, China}
\author{Fu-Zhou Chen}
\affiliation{School of Physical Science and Technology $\&$ Lanzhou Center for Theoretical Physics, Lanzhou University, Lanzhou 730000, China}
\affiliation{Key Laboratory of Quantum Theory and Applications of MoE $\&$ Key Laboratory of Theoretical Physics of Gansu Province, Lanzhou University, Lanzhou 730000, China}
\author{Chen Cheng}
\affiliation{School of Physical Science and Technology $\&$ Lanzhou Center for Theoretical Physics, Lanzhou University, Lanzhou 730000, China}
\affiliation{Key Laboratory of Quantum Theory and Applications of MoE $\&$ Key Laboratory of Theoretical Physics of Gansu Province, Lanzhou University, Lanzhou 730000, China}
\author{Hong-Gang Luo}
\email{luohg@lzu.edu.cn}
\affiliation{School of Physical Science and Technology $\&$ Lanzhou Center for Theoretical Physics, Lanzhou University, Lanzhou 730000, China}
\affiliation{Key Laboratory of Quantum Theory and Applications of MoE $\&$ Key Laboratory of Theoretical Physics of Gansu Province, Lanzhou University, Lanzhou 730000, China}

\begin{abstract}
The frustrated spin-$1/2$ $J_1-J_2$ Heisenberg model on the square lattice has been extensively studied since 1988 because of its close relationship to the high-temperature superconductivity in cuprates and more importantly involved novel phase of matter in its own right, namely, quantum spin liquid (QSL), one of hot topics in condensed matter physics in recent years. However, the phase diagram of the model, particularly in the maximally frustrated regime $J_2/J_1 \sim 0.5$, is quite controversial, and more seriously the nature of the QSL is not clear at all. Here we provide a pattern picture, on one hand, to show explicitly how the system evolves from the N{\'e}el antiferromagnetic (AFM) state at small $J_2$ to the striped AFM one at large $J_2$; on the other hand, to uncover the nature of the QSL if it exists in the intermediate $J_2$ coupling regime. For simplicity, we show our results by taking the square lattice $L=L_x \times L_y$ with size $L_x=L_y=4$ here and periodic boundary condition is considered, and furthermore, exact diagonalization is employed to confirm the correctness of our picture. Our results indicate that the highly frustration regime is characterized by diagonal two-domain, while the N{\'e}el AFM state has a diagonal single-domain and the striped AFM state shows itself as a diagonal four-domain, namely, completely diagonal antiferromagnetic order, in the present case. Increasing the system size, the number of the diagonal domains increases correspondingly, but the diagonal single-domain for the N{\'e}el AFM state and the diagonal $L_{x(y)}$-domain for the striped AFM state remain unchanged. Our results shed light on the understanding of the QSL.
\end{abstract}
\maketitle

\section{Introduction}
Fifty years have passed \cite{Kivelson2023} since the original concept of quantum spin liquid (QSL) has been introduced by Anderson in 1973, called a `resonating valence bond' state formed by firstly pairing spins into singlets and then making a quantum superposition of these singlets at the wavefunction level \cite{Anderson1973}. The physical importance of the QSL may come from two aspects. The first is that the QSL, in its own right, is a kind of new phase of matter, showing no magnetic orders even in the limit of zero temperature but involving many exotic properties such as non-local excitations, topological nature, and so on. For details, one can refer to some reviews \cite{Balents2010, Savary2017, Zhou2017, Wen2017, Wen2019, Broholm2020}. This is obviously beyond the conventional Landau's paradigm on understanding the phases of matter and their transitions \cite{Kalmeyer1987, Kivelson1987, Senthil2000, Knolle2019}. These exotic properties of QSL are believed to hold great potential for quantum communication and computation \cite{Kitaev2003, Kitaev2006, Nayak2008, Lahtinen2017}. On the other hand, it was suggested by Anderson in 1987 that high-temperature superconductivity firstly found in cuprates in 1986 \cite{Bednorz1986} might emerge by doping a QSL \cite{Anderson1987, Inui1988, Manousakis1991}. This is another motivation to explore the nature of QSL because the mechanism of the high-temperature superconductivity remians still open, despite the fact that more and more high-temperature superconducting materials have been found. 

However, it indicates that it is actually difficult to make a \textit{positive} answer to the question: what is a QSL? In practice, it usually makes a negative way to identify possible QSLs when neglecting some storied examples (see below). Theoretically, it is popularly believed that a QSL should have a ground state with highly entangled nature, namely, the state is a quantum superposition one, which \textit{cannot} be decomposed as a product state by any way through changes of local basis. Although what the highly entangled state looks like or what possible QSLs are have been shown clearly by Kitaev's toric code model \cite{Kitaev2003} and Kitaev's honeycomb model \cite{Kitaev2006}, respectively, there are no realistic materials to follow completely the physics described by these two models till today. The honeycomb iridate materials were believed to be able to realize the Kitaev-type interaction between neighboring spins \cite{Jackeli2009}. Unfortunately, an additional Heisenberg term is unavoidable due to the direct Ir-Ir overlap and furthermore, this kind of materials are magnetic in nature below certain temperatures \cite{Liu2011,Choi2012}, which clearly excludes the QSL in these materials. This brings a most popular means to identify candidate materials (models) which might host QSL physics that is to demonstrate experimentally (numerically) the lack of magnetic ordering. This is another negative way to search possible QSLs: what the spins do \textit{not} do. Actually, this is the key difficulty in the study of the QSL physics up to now. 

In this work we provide a \textit{positive} attempt to show what the spins do for a given system that is assumed to host a QSL phase. The frustrated spin-$1/2$ $J_1$-$J_2$ antiferromagnetic Heisenberg model on a square lattice is such an example ($J_1$ and $J_2$ are the nearest- and next-nearest-neighbor exchange couplings, respectively). Initial interest in this model was brought from its connection to the high-temperature superconductivity in cuprates \cite{Anderson1987, Inui1988, Wen1989, Manousakis1991}. It is well-known that this frustrated model is in the N{\'e}el AFM state if $J_2$ is small; otherwise, the model is in the striped AFM phase. But the intermediate $J_2$ regime around $J_2 \sim J_1/2$ is still enigmatic \cite{Schulz1992, Schulz1996, Singh1999, Capriotti2000, Mambrini2006, Beach2009, Richter2010, Yu2012, Jiang2012, Wang2013, Hu2013, Gong2014, Doretto2014, Haghshenas2018, Wang2018, Ferrari2020, Nomura2021, Liu2022}. Different candidate ground states were proposed by diffferent kinds of methods, including a plaquette valence-bond solid (VBS) state \cite{Capriotti2000, Mambrini2006, Yu2012, Gong2014, Doretto2014}, a columnar VBS state \cite{Singh1999, Haghshenas2018}, a gapless QSL state \cite{Hu2013, Wang2013, Wang2018, Ferrari2020, Nomura2021, Liu2022} and a gapped QSL state \cite{Jiang2012}. Even the same method could lead to different conclusions. For example, by using density matrix renormalization group three kinds of phases have been addressed, namely, gapped spin liquid \cite{Jiang2012}, plaquette VBS \cite{Gong2014} or gapless spin liquid \cite{Wang2018}. Essential difficulties may come from inadequate computational resources and more importantly unclear definition of QSL or indirect evidences such as orders or correlation function behaviors extracted from finite size scaling. Here we employ a pattern picture proposed in our previous works \cite{Yang2022c,Yang2023a,Yang2023b,Yang2023c} to tackle the frustrated spin-$1/2$ $J_1$-$J_2$ model. Rather successful application of this picture to the one-dimensional (1D) frustrated spin model \cite{Yang2023c} encourages us to consider this two-dimensional (2D) version of the frustration models and to explore the nature of possible QSL physics in this model. 

For the sake of clarity, we take lattice size $L=L_x\times L_y$($L_x=L_y=4$), which is also treated readily with numerical exact diagonalization (ED). The cases with larger lattice sizes are left for the future study. In what follows, periodic boundary condition (PBC) is taken. All spectra of model are readily obtained and the ground state and low-lying excited states can be analyzed in detail. As done previously \cite{Yang2023c}, we firstly diagonalize the Hamiltonian in operator space consisting of spin components at each lattice site, thus obtain six kinds of patterns with high degeneracy (given later). We define domains/kinks along the diagonal orientation for the square lattice, reminiscent of those in the 1D case. The results show that the patterns $\lambda_1$-$\lambda_3$ have diagonal single-domain which is identified as N{\'e}el AFM pattern and the patterns $\lambda_4$-$\lambda_{15}$ have diagonal two-domain. The patterns $\lambda_{16}$-$\lambda_{21}$ have diagonal four-domain, identified as striped AFM pattern. With these patterns at hand, we show explicitly how the ground and the first excited states evolve with increasing $J_2$: for low $J_2$, the patterns $\lambda_1$-$\lambda_3$ dominate over the others, and then fade away rapidly; at the same time, the patterns $\lambda_4$-$\lambda_{15}$ begin to come into play in the intermediate $J_2$ coupling regime and are suppressed once $J_2$ increases further; Finally, the patterns $\lambda_{16}$-$\lambda_{21}$ become dominant in the large $J_2$ regime, and the system enters into the striped AFM phase. Increasing the lattice size, there are more diagonal domains such as two-, four-, six-, $\cdots$, $(L_{x(y)}-2)$-domain in the intermediate nonmagnetic regime to come successively into play, and finally the diagonal $L_{x(y)}$-domain is identified as the striped AFM phase. In the following we provide the detail to confirm our statement.

\begin{figure*}[tbp]
\begin{center}
\includegraphics[width =1.8 \columnwidth]{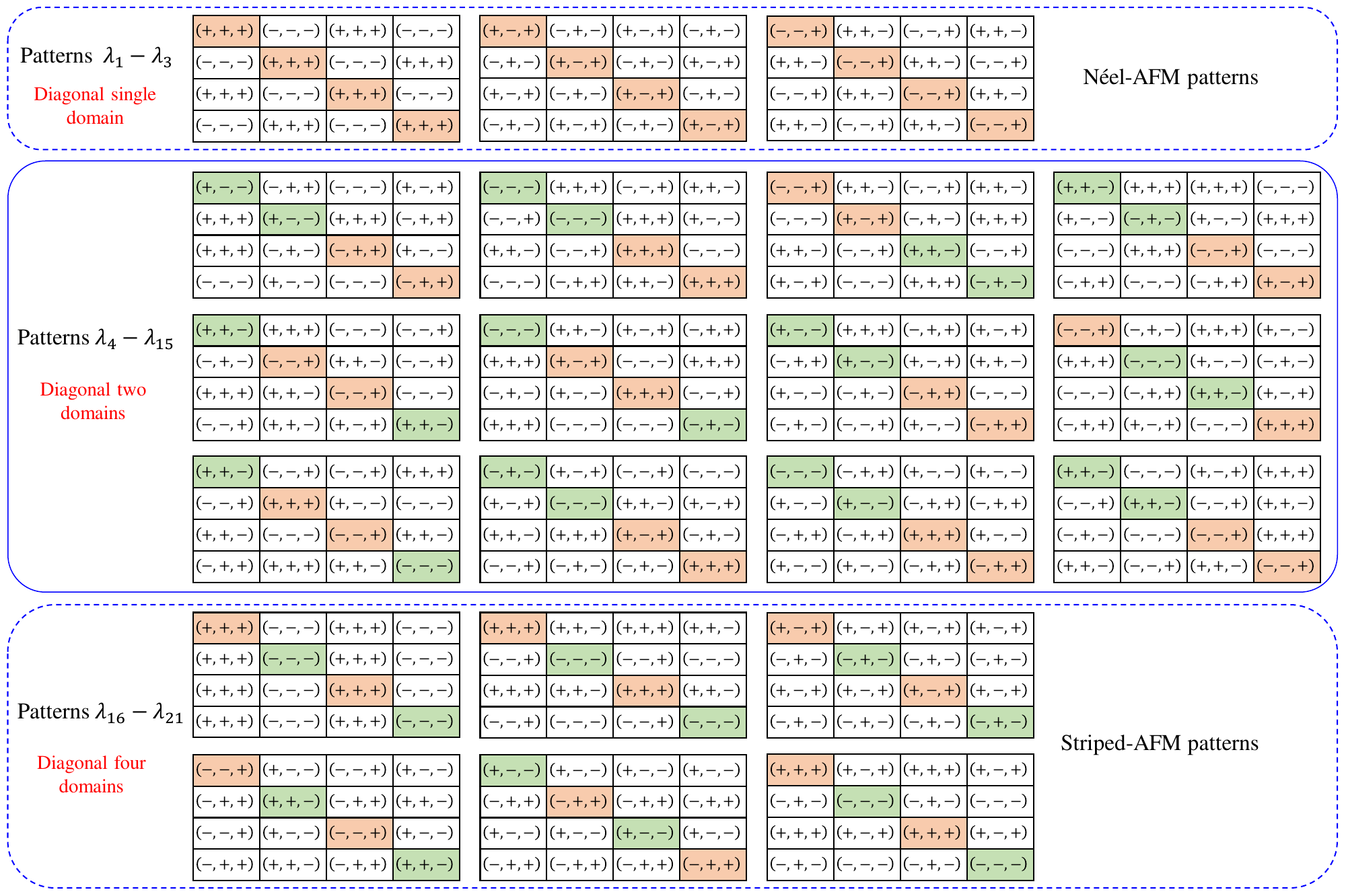}
\caption{The patterns and their relative phases obtained by the first diagonalization, marked by the single-body operators Eq. \eqref{j1j23b} with $(\pm,\pm,\pm)$ denoting the signs of $(u_{n,3i-2},u_{n,3i-1},u_{n,3i})$ for the square lattice $J_1$-$J_2$ model with $L=4\times4$ under PBC. The four diagonal sites are marked by orange and green. Orange sites represent $(\pm,\pm,+)$ and green sites represent $(\pm,\pm,-)$. In addition, the patterns' eigenfunctions are free of a total phase factor $e^{i\pi}$ but their relative phases remain fixed.} \label{fig1}
\end{center}
\end{figure*}

\section{Model and Method}
The Hamiltonian of the frustrated $J_{1}$-$J_{2}$ Heisenberg antiferromagnet reads
\begin{equation}
\hat{H}' = J_1\sum_{\langle i,j\rangle}\boldsymbol{\hat S_i \cdot \hat S_{j}} + J'_2\sum_{\langle\langle i,j\rangle\rangle}\boldsymbol{\hat S_i \cdot \hat S_{j}}, \label{j1j20}
\end{equation}
where $\langle i,j\rangle$ and $\langle\langle i,j\rangle\rangle$ represent the nearest- and next nearest-neighbor exchange interactions of the spins located at sites $i$ and $j$, respectively. In what follows, we take $J_1$ as units of energy and $J_2 = \frac{J'_2}{J_1}$, representing a dimensionless frustration parameter. Thus Eq. (\ref{j1j20}) is reformulated as $\hat{H}'= \frac{J_1}{2} \hat{H}$, where 
\begin{small}
\begin{equation}
\hat{H} = \sum_{\langle i,j\rangle} \left(\boldsymbol{\hat S_i \cdot \hat S_{j} + \hat S_j \cdot \hat S_{i}}\right) + J_2 \sum_{\langle\langle i,j\rangle\rangle} \left(\boldsymbol{\hat S_i \cdot \hat S_{j} + \hat S_j \cdot \hat S_i}\right). \label{j1j21}
\end{equation}
\end{small}
Furthermore, Eq. \eqref{j1j21} can be written equivalently as a $3L\times 3L$ matrix
\begin{eqnarray}
&&\hat{H} = \left(
\begin{array}{cccccccccc}
\hat S^x_1& i\hat S^y_1& \hat S^z_1& \hat S^x_2& i\hat S^y_2& \hat S^z_2 & \cdots & \hat S^x_{L}& i\hat S^y_{L}& \hat S^z_{L}
\end{array}
\right)\nonumber\\
&&\times
\left(
\begin{array}{cccccccccc}
0 &0 &0 &1 & 0 &0 &\cdots &J_2 &0 &0 \\
0 &0 &0 &0 & 1 &0 &\cdots &0 &J_2 &0 \\
0 &0 &0 &0 &0 &1 &\cdots &0 &0 &J_2 \\
1 &0 &0 &0 &0&0 &\cdots &0 &0 &0 \\
0 &1 &0 &0 &0 &0 &\cdots &0 &0 &0 \\
0 &0 &1 &0 &0 &0 &\cdots &0 &0 &0 \\
\vdots &\vdots &\vdots &\vdots &\vdots &\vdots &\ddots &\vdots &\vdots &\vdots \\
J_2 &0 &0 &0 &0&0&\cdots &0 &0 &0 \\
0 &J_2 &0 &0 &0&0 &\cdots &0 &0 &0 \\
0 &0 &J_2 &0 &0&0 &\cdots &0 &0 &0
\end{array}
\right)\times \label{j1j22}\\
&&\left(
\begin{array}{cccccccccc}
\hat S^x_1& -i\hat S^y_1& \hat S^z_1& \hat S^x_2& -i\hat S^y_2& \hat S^z_2& \cdots & \hat S^x_{L}& -i\hat S^y_{L}& \hat S^z_{L}
\end{array}
\right)^T, \nonumber
\end{eqnarray}
where the superscript $T$ denotes transpose. This matrix can be diagonalized to obtain eigenvalues and corresponding eigenfunctions $\{\lambda_n, u_n\}  (n = 1, 2, \cdots, 3L)$, which define the patterns marked by $\lambda_n$. Thus the $J_1$-$J_2$ Hamiltonian is rewritten as
\begin{equation}
\hat{H} = \sum_{n=1}^{3L} \lambda_n \hat{A}^\dagger_n \hat{A}_n, \label{j1j23a} 
\end{equation}
where each pattern $\lambda_n$ composes of single-body operators
\begin{equation}
\hat{A}_n = \sum_{i=1}^{L}\left[u_{n,3i-2} \hat S^x_{i} + u_{n,3i-1} (-i\hat S^y_{i}) + u_{n,3i} \hat S^z_{i}\right]. \label{j1j23b}
\end{equation}
The validity of Eq. (\ref{j1j23a}) can be confirmed by inserting into the complete basis $|\{S^z_i\}\rangle (i = 1, 2,\cdots, L)$ with $\hat{S}^z_i |\{S^z_i\}\rangle = \pm_i(\uparrow,\downarrow) |\{S^z_i\}\rangle$, as done in Refs.\cite{Yang2022c, Yang2023a, Yang2023b,Yang2023c}.

\begin{figure}[tbp]
\begin{center}
\includegraphics[width =0.8 \columnwidth]{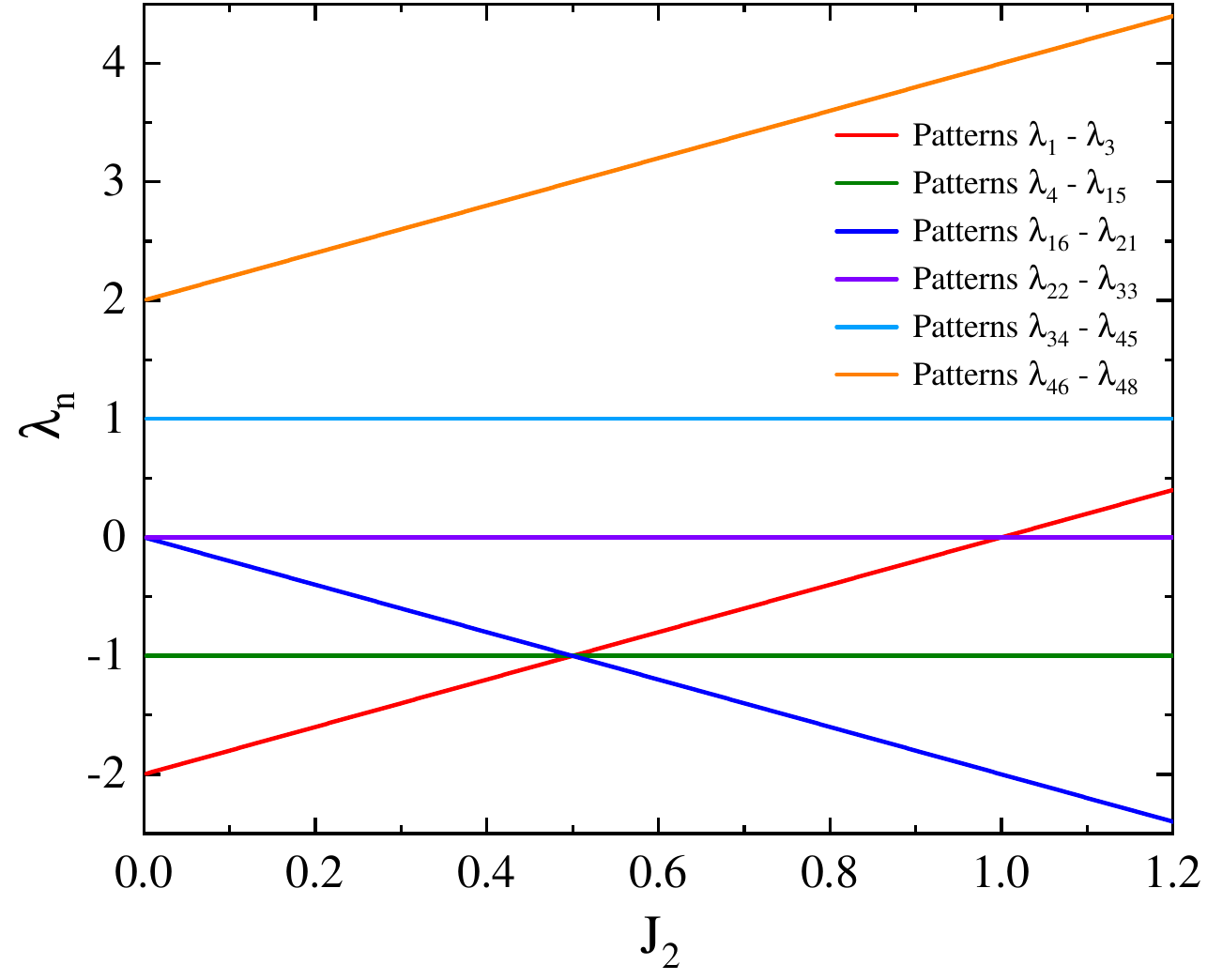}
\caption{The eigenenergies of the patterns as functions of the frustrated parameter $J_2$ for $L=4 \times 4$ lattice size. Here and hereafter the step of $J_2$ is taken as 0.003.}\label{fig2}
\end{center}
\end{figure}

\section{Patterns' Information} 
Firstly we explore the properties of the patterns marked by relative signs of eigenfunctions $(u_{n,3i-2},u_{n,3i-1},u_{n,3i})$, shown in Fig. \ref{fig1} and their eigenvalues $\lambda_n$ as functions of $J_2$, given in Fig. \ref{fig2}, from which six kinds of patterns with different degeneracy exist. The first kind of pattern (red line) is threefold-degeneracy which is named the patterns $\lambda_1$-$\lambda_3$. The second one (green line) is twelvefold-degeneracy, named the patterns $\lambda_4$-$\lambda_{15}$ and the third one (blue line) is sixfold-degeneracy, named the patterns $\lambda_{16}$-$ \lambda_{21}$. These three kinds of patterns have negative eigenenergies in the most region of the $J_2$ coupling we consider, thus are important to the ground state and the low-lying excited states we are interested in. The remainder is either zero (the patterns $\lambda_{22}$-$\lambda_{33}$) or positive eigenenergies (the patterns $\lambda_{34}$-$\lambda_{45}$ and $\lambda_{46}$-$\lambda_{48}$), which have zero or less contribution to the cases we consider here. Therefore, we only list the details in Fig. \ref{fig1} for first three kinds of patterns. 

Here we borrow the concept of magnetic domains/kinks in the 1D spin chain by checking the diagonal direction of square lattice and focus on the domain numbers (due to PBC used, the concept of diagonal domains/kinks behaves well for arbitrary diagonal direction). For each site, there are three spin components $S^x_i, -iS^y_i$ and $S^z_i$ describing the spin state, and we assign the spin $z$-component to express the spin state and thus other two components represent the dynamical information of the spin at that site. Therefore, we mark the spin states according to the coefficients of $S^z_i$, i.e. the signs of $u_{n,3i}$. Obviously, the patterns $\lambda_1$-$\lambda_3$ have a character of diagonal single-domain and the nearest-neighbor lattices are completely antiferromagnetic for all three components. Thus, the patterns $\lambda_1$-$\lambda_3$ are identified as the N{\'e}el AFM state. The patterns $\lambda_4$-$\lambda_{15}$ have a feature of diagonal two-domain, as marked by different colors. The patterns $\lambda_{16}$-$\lambda_{21}$ behave as diagonal four-domain, marked by alternating colors and a typical striped AFM behavior is obtained. Therefore, the patterns $\lambda_{16}$-$\lambda_{21}$ are identified as striped AFM state.

After the characteristic features of the patterns have been identified, it is interesting to check the corresponding eigenenergies of these patterns, shown in Fig. \ref{fig2}. Obviously, at small $J_2$, the patterns $\lambda_1$-$\lambda_3$ have lower energies, thus dominate this regime. As a result, this regime is the N{\'e}el AFM phase in nature; at large $J_2$, the patterns $\lambda_{16}$-$\lambda_{21}$ have lower eignenergies, which thus dominate this regime by the striped AFM phase; in the intermediate $J_2$, i. e. the maximally frustration regime $J_2 \sim J_1/2$, the situation is lightly complicated since this is the regime with a classical high-degenerate point and three kinds of lower-energies patterns compete with each other. In this case, a more careful calculation is needed, as given in the following. 

\begin{figure}[tbp]
\begin{center}
\includegraphics[width = \columnwidth]{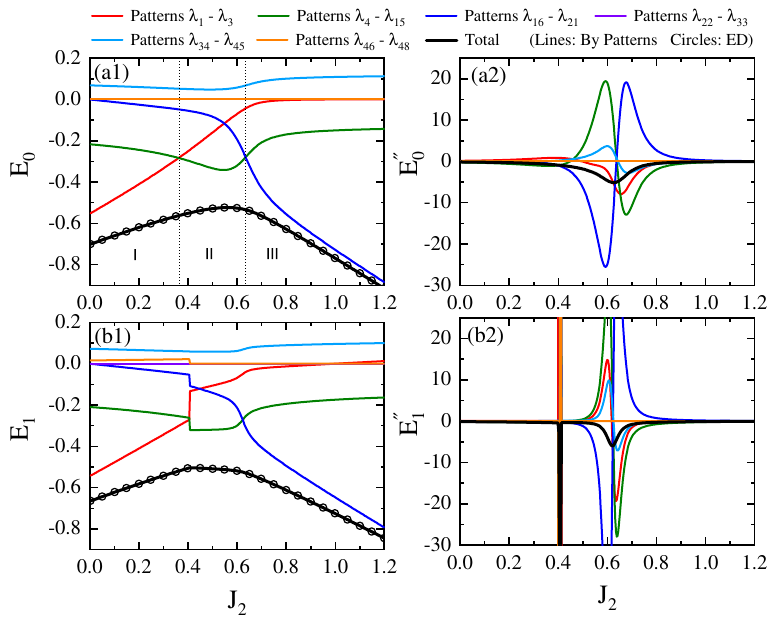}
\caption{(a1) $\&$ (b1) The ground state and the first excited state energies as functions of the frustration parameter $J_2$ (thick black solid ines) and their pattern dissections (thin color solid lines). Direct numerical ED (circles) results are also presented to check the validity of the pattern picture. The vertical dashed-lines in (a1) marks two cross points of $J_{2,c1} \approx 0.363$ and $J_{2,c2} \approx 0.635$, respectively, which divide the frustration parameter $J_2$ into three regions: I, the N{\'e}el AFM; II, the diagonal two-domain dominant; and III, the striped AFM. (a2) $\&$ (b2) The second derivatives of the corresponding energy levels (thick black solid lines) and their pattern dissections (thin color solid lines). }\label{fig3}
\end{center}
\end{figure}

Firstly, one readily obtains the matrix $\left[\hat{A}_n\right]_{\{S^z_i\},\{S^z_i\}^{\prime}} = \langle\{S^z_i\}|\hat{A}_n|\{S^z_i\}^{\prime}\rangle$ and then Eq. (\ref{j1j23a}) can be solved by diagonalizing the matrix with elements
\begin{eqnarray}
&& \langle \{S^z_i\}|\hat{H}|\{S^z_i\}^{\prime}\rangle = \sum_{n=1}^{3L} \lambda_n \nonumber\\
&& \hspace{1cm}\times \sum_{\{S^z_i\}^{\prime\prime}} \left[\hat{A}^\dagger_n\right]_{\{S^z_i\},\{S^z_i\}^{\prime\prime}}\left[\hat{A}_n\right]_{\{S^z_i\}^{\prime\prime},\{S^z_i\}^{\prime}}.\label{Ising4}
\end{eqnarray}
Figure \ref{fig3} (a1) $\&$ (b1) present the results for the ground state and the first excited state energies as functions of $J_2$, as shown by thick black solid lines, respectively. In order to confirm the validity of the pattern formulation, the results of direct ED have also presented for comparison, shown as circles. The exact agreement between them from weak to strong $J_2$ regimes is noticed, which is not surprising since no any approximation has been introduced. 

\section{Evolution from N{\'e}el to Striped AFM States} 
The dissection of the ground state energy is plotted in Fig. \ref{fig3} (a1), which can be roughly divided into three regions with increasing $J_2$. I, the N{\'e}el AFM phase, where the patterns $\lambda_1$-$\lambda_3$ dominate over others, but the patterns $\lambda_4$-$\lambda_{15}$ have a remarkable contribution, and even become more and more important. Obviously, this region is somehow different to the standard N{\'e}el AFM state due to the presence of the frustration, but we still call it the N{\'e}el AFM phase; II, the region that the patterns $\lambda_{4}$-$\lambda_{15}$ dominate over others. In this region, the contribution from the patterns $\lambda_1$-$\lambda_3$ fades rapidly away and that from the patterns $\lambda_{16}$-$\lambda_{21}$ still grows up. In this case, this region is mainly characterized by the patterns $\lambda_{4}$-$\lambda_{15}$, which have diagonal two-domain in nature, as shown in Fig. \ref{fig1}. The cross point is about $J_{2,c1} \approx 0.363$; III, the striped AFM phase beginning about at $J_{2,c2}\approx 0.635$. In this region, the patterns $\lambda_{16}$-$\lambda_{21}$ dominate over others. These patterns have diagonal four-domain, which are completely antiferromagnetic order in the diagonal direction. As a result, the spins form striped structure characterized by the striped AFM state. Still, the contribution from the patterns $\lambda_{4}$-$\lambda_{15}$ is prominent, but that from the patterns $\lambda_1$-$\lambda_3$ becomes almost zero. In addition, the patterns $\lambda_{34}$-$\lambda_{45}$ have a minor positive contribution due to the interplay between the quantum fluctuation and the magnetic exchange interactions. In addition, other patterns have no contributions to the ground state energy due to either zero pattern energy such as the patterns $\lambda_{22}$-$\lambda_{33}$ or too high pattern energy such as the patterns $\lambda_{46}$-$\lambda_{48}$.
 
\begin{figure}[tbp]
\begin{center}
\includegraphics[width = \columnwidth]{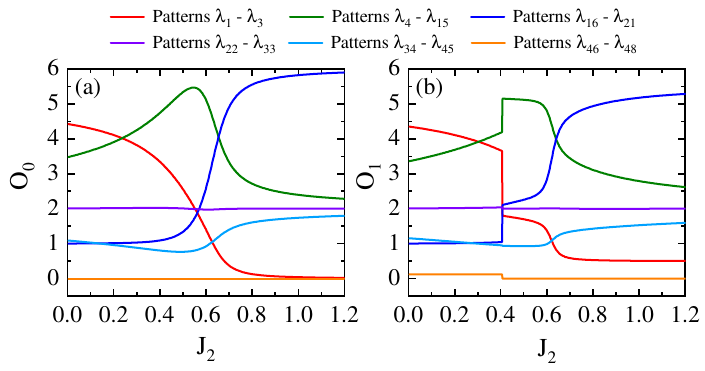}
\caption{(a) The patterns' occupancy in the ground state and (b) the first excited state with $L=4 \times 4$ as functions of $J_2$. }\label{fig4}
\end{center}
\end{figure}

The above analysis indicates that the pattern energies have a rough correspondence to the actual contributions of the corresponding patterns to the ground state energy, which is specially ture in the small and large $J_2$ regimes since the patterns $\lambda_1$-$\lambda_3$ and the patterns $\lambda_{16}$-$\lambda_{21}$ have apparently lower eigenenergies than the others, respectively. However, in the maximal frustration regime, their eigenenergies are comparable, and in particular, are degenerate exactly at $J_2 = J_1/2$, a classical high-degenerate point. The contributions of the patterns to the ground state are mainly determined by the patterns occupancy calculated by $O_0=\langle\Psi_0|\hat{A}^\dagger_n \hat{A}_n|\Psi_0\rangle$ where $|\Psi_0\rangle$ is the ground state wavefunction. Therefore, the distinguished contribution of the patterns $\lambda_{4}$-$\lambda_{15}$ to the ground state comes from that they have large occupancy, as shown in Fig. \ref{fig4} (a). Thus, the physics of this region is dominated by a diagonal two-domain structure. If a QSL is assumed to exist in this region, the diagonal domain structure is a natural character of the QSL. Surprisingly, these two cross points of $J_{2,c1}$ and $J_{2,c2}$ are roughly consistent with those reported in the literature \cite{Richter2010}, irrespective of the small lattice size we consider here. 

The first excited state has also been analyzed by the same way, as shown in Fig. \ref{fig3} (b1) and Fig. \ref{fig4} (b). The contributions of different patterns have similar behaviors, but the first excited state has a first-order phase transition at $J_2 = 0.4$. The phase transition around $J_2 \sim 0.6$ behave more like a continuous one. Figs. \ref{fig3} (a2) and (b2) provide the second derivative of the ground state and the first excited state energies, which confirm the above statement. 

\section{Summary and Discussion}
We use a pattern picture to solve the frustrated spin-1/2 $J_1$-$J_2$ antiferromagnetic Heisenberg model on the square lattice with a small size of $L = 4\times 4$, which is confirmed by ED result. Our aim is not to solve analytically or numerically this model in a traditional sense, but to explore what the spins do in the maximally frustrated regime, a quite controversial regime, even in the case of large lattice size. On the other hand, this regime is involved an issue of the nature of the QSL if it exists. Our work arrives at this aim, in which a diagonal two-domain structure characterizes this regime, which shed light on the understanding of the QSL in a \textit{positive} sense. 

Due to small lattice size, the maximally frustrated regime only dominates by a diagonal two-domain structure. Increasing the lattice size, one can image that there are more domain structures such as diagonal four-, six-, $\cdots$, $(L_{x(y)}-2)$-domain to occur, and they will dominate successively the ground state with increasing $J_2$. This will be testified for larger lattice size in the future study. In addition, it is also interesting to check other lattice systems like the triangular, honeycomb, Kagom{\'e}, and more, which are useful to further check the nature of the QSL. 

\section{Acknowledgments}
The work is partly supported by the National Key Research and Development Program of China (Grant No. 2022YFA1402704) and the programs for NSFC of China (Grant No. 11834005, Grant No. 12174167, Grant No. 12247101).




%

\end{document}